# Electro-Chemo-Mechanical Properties in Nanostructured Ca-doped Ceria (CDC) by Field Assisted Sintering

Ahsanul Kabir[a*], Haiwu Zhang[a], Sofie Colding-Jørgensen[a], Simone Santucci[a], Sebastian Molin[b], Vincenzo Esposito[a*]

[a]Department of Energy Conversion and Storage, Technical University of Denmark, 2800 Kgs. Lyngby, Denmark

[b]Advanced Materials Center, Faculty of Electronics, Telecommunications, and Informatics, Gdańsk University of Technology, ul. G. Narutowicza 11/12, 80-233 Gdańsk

*Corresponding Authors: E-mail: ahsk@dtu.dk, vies@dtu.dk

Abstract

Recent investigations have shown that highly oxygen defective cerium oxides generate non-classical electrostriction that is superior to lead-based ferroelectrics. In this work, we report the effect of field-assisted spark plasma sintering (SPS) on electro-chemo-mechanical properties on Ca-doped ceria (CDC). Nanometric powders of *ca.*10 nm are rapidly consolidated to form polycrystalline nanostructures with a high degree of crystalline disorder. Remarkably, the resultant material demonstrates a large electromechanical strain without a frequency-related relaxation effect. We conclude that electromechanical activity in CDC materials strictly depends on the $Ca^{2+}$-$V_O^{\cdot\cdot}$ interaction, while disorder at the crystalline boundaries has a minor effect.

Keywords: Calcium doped ceria (CDC), Spark plasma sintering, Nanostructures, Ionic conductivity, Electrostriction

Rare-earth doped cerium oxide is widely used in electroceramics as a solid electrolyte and electrode material for solid oxide fuel cells (SOFCs), catalysts, gas sensors, gas separation membranes, *etc.* [1][2][3]. In addition to these well-known applications, an unusual electromechanical activity, namely electrostriction effect, is recently demonstrated in both thin films and bulk polycrystalline form of Gd-doped ceria (GDC) [4][5][6][7]. The electrostrictive strain coefficient ($M_e$) is reported between ~$10^{-17}$-$10^{-18}$ (m/V)$^2$ depending on the operative frequency of the electric field. Such a large response is unprecedented, as ceria has a centrosymmetric crystal structure with a relatively low dielectric constant ($\varepsilon_r^{GDC} \approx 30$) [8], signifying different electrostriction mechanisms with respect to those currently known [9] are in action, i.e. non-classical electrostriction [10]. Lubomirsky and co-workers have demonstrated that this effect is driven by the presence of oxygen vacancies ($V_O^{\cdot\cdot}$) in the crystal lattice [11][12]. The oxygen vacancies locally distort the unit cell by creating electroactive elastic $Ce_{Ce}$-$V_O^{\cdot\cdot}$ dipoles. Upon interaction with the electric field, defects "rattle", leading to straining and relaxing of the structure [13]. Accordingly, significant macroscopic electromechanical activity is created. To date, electrostriction in cerium oxides is only reported in GDC compositions. In particular, the role of other parameters, for instance, dopant types, mass-diffusion mechanism, and processing parameters on the electromechanical property are not yet thoroughly examined.

In this work, we investigate the electro-chemo-mechanical properties of 5 mol% calcium-doped ceria (CDC), consolidated by field-assisted spark plasma sintering (SPS). The use of SPS allows nanostructuring the material, increasing the density of the ion-blocking barriers. When compared to the Gd$^{3+}$ doping in GDC, divalent calcium dopant Ca$^{2+}$ enhances the polarization at the lattice *via* a stronger Ca$^{2+}$-$V_O^{\cdot\cdot}$ attraction due to the 1:1 ratio of the charge between Ca$^{2+}$ and the oxygen vacancy. Additionally, the ionic radius of Ca$^{2+}$ (1.12 Å) is larger than Gd$^{3+}$ (1.05 Å) in octahedral coordination, resulting in a relatively high elastic strain (~15.5 %) in the CDC lattice The combined effect of both the columbic and elastic interaction is expected to impact directly the inherent properties in the material. Finally, the results are compared with one conventionally sintered microcrystalline CDC material and one 10 mol% GDC as an equivalent reference in terms of nominal oxygen defects concentration, as reported in Ref. [6].

Nano-sized 5 mol% calcium doped ceria (CDC) powders were prepared by the co-precipitation method, as described elsewhere [6]. The starting powders were consolidated by the SPS method (Dr. Sinter Lab 515S) at 980 °C, with a uniaxial pressure of 50 MPa and 5 min dwelling. The as-sintered sample was then reoxidized at 800 °C for 1 hour, to maintain equilibrium oxygen vacancy concentration. For comparison, one SPS sample was further post-annealed at 1450 °C in the air for 10 hours. The morphology and average particle size of the powders were analyzed by a transmission electron microscope (TEM, JEOL 2100, $LaB_6$) under 200 kV. The powder sample was mixed with ethanol, sonicated for 10 s and directly drop cast on to a TEM grid. The crystallographic phase composition was verified by the X-ray diffraction technique (XRD, Bruker D8, CuKα). A high-resolution scanning electron microscope (SEM, Zeiss Merlin) was used to characterize the microstructure. The grain is assessed by the linear intercept method accounting for more than 100 grains. The electrochemical properties were investigated using a Solarton (1260) at a temperature between 300-600 °C in the frequency range of 10 Hz-10 MHz at an amplitude signal of 100 mV. Non-volatile silver paste (SPI) was brushed onto the parallel surface and dried at 600 °C for 15 min. The resultant data were fitted employing an equivalent circuit model and analyzed by the ZView software shareware version. The electromechanical strain measurement was performed with a single beam laser interferometer (SIOS NA Analyzer) coupled with a lock-in amplifier. The sample was coated with a gold (Au) electrode with a thickness of ~50-80 nm.

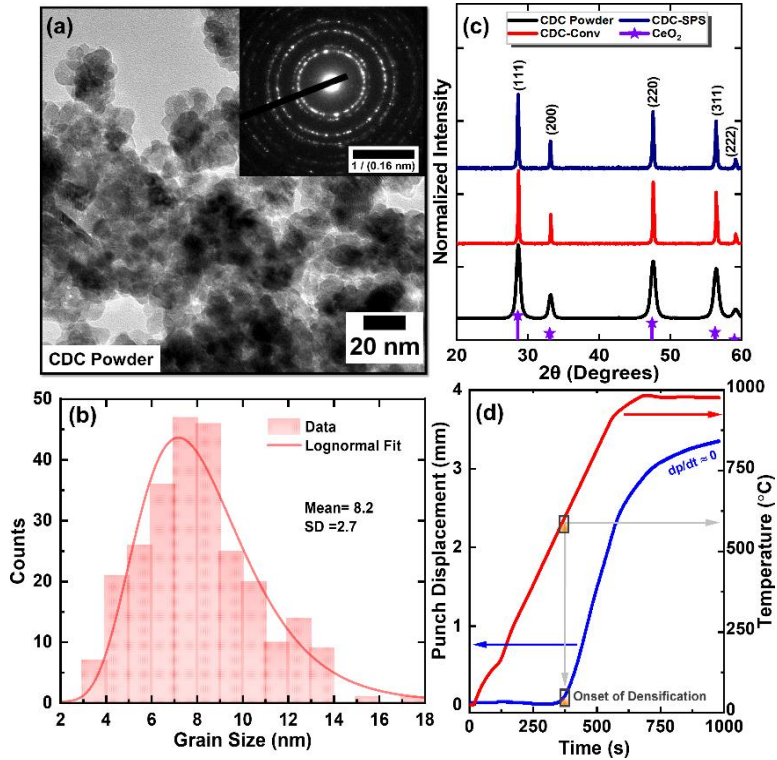

**Figure 1:** (a) A bright-field TEM image of the CDC powders, inset shows corresponding selected area diffraction (SAED) pattern, (b) Particle size distribution of the powders, (c) X-ray diffraction (XRD) patterns of CDC powders and sintered pellets, indexed with pure ceria (ICSD # 251473), (d) Densification profile in SPS, at 980 °C for 50 MPa for 5 min dwelling.

The particle size analysis and morphology of the starting CDC powders are illustrated in **Fig. 1.a** and **Fig. 1.b**. As observed, powders are loosely agglomerated and homogeneously distributed. The particle has a typical spherical shape, having an average particle size of about 8.2 ± 2.7 nm. The inset of **Fig. 1.a** represents a characteristic selected area electron diffraction (SAED) pattern, showing a cubic fluorite structure of ceria. A similar crystallographic structure is also illustrated by the X-ray diffraction (XRD) for both powders and sintered pellets (**Fig. 1c**). Considering the resolution limit of XRD, no crystal planes other than fluorite are observed. The experimental density of the pellets was measured by the Archimedes method and is above 95% of theoretical density. The sintering profile of the SPS (**Fig. 1d**) demonstrates that the densification starts around ~600 °C and almost finishes at the holding segment when the slope of the punch displacement against time curve tends to become zero.

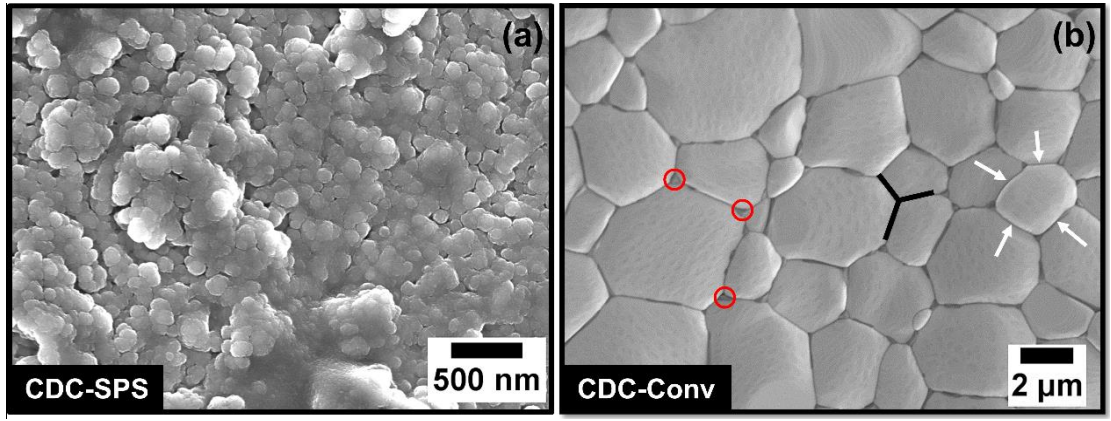

**Figure 2:** The SEM micrographs of CDC pellets, sintered in (a) SPS at 980 °C for 5 min, (b) air at 1450 °C for 10 h. The images are taken in secondary electron modes under 4 kV.

The experimental density of the samples is measured via the Archimedes method in deionized water and above 95% of their theoretical density. **Fig. 2** illustrates the microstructural evolution of the CDC samples. As noticed, the CDC-SPS sample exhibits nanostructured polygonal grains with an average size of ~150 ± 20 nm. These grains are non-relaxed, with a presence of some residual non-interconnected porosity. The restriction of grain growth in SPS is attributed to fast heating rates leading to localized large sintering temperature at the particle necks [14], eventually circumventing grain coarsening associated initial stage of sintering i.e. surface diffusion. On the other hand, the CDC-Conv sample sintered at 1450 °C, registers significant grain growth, having a grain size above 4 µm. These grains are mostly relaxed, have equilibrium configuration at the triple point and small residual grain boundary curvature (black and white marks in **Fig. 2.b**). Moreover, the thin line at the grain-to-grain contact and triple point (red circle) probably indicate enrichment of calcium. Calcium segregation in Ca-doped ceria was largely observed before and it is generally attributed to solute drag mechanisms during long thermal treatments [15][16].

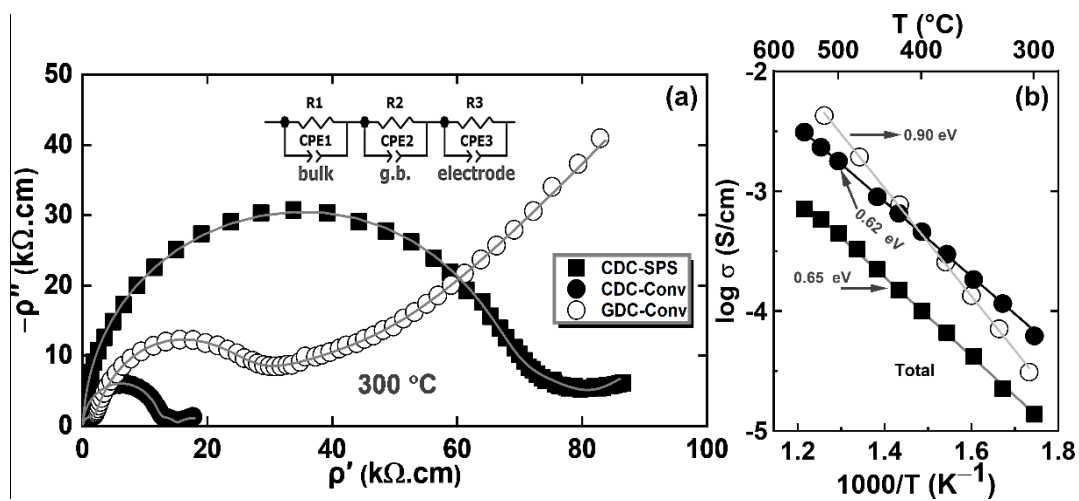

**Figure 3:** (a) Representation of the geometry normalized Nyquist plots (ρ' vs ρ") at 300 °C, measured in air with Ag electrode (b) Arrhenius plot for the estimation of the temperature-dependent total electrical conductivity. The data of the GDC-Conv is taken after Ref. [6].

The electrochemical properties of the samples are represented in **Fig. 3a** as Nyquist plot formalism from data collected by electrochemical impedance spectroscopy (EIS). Such a methodology allows separating capacitive and ohmic properties in the materials related to electrically charged species (i.e. ionic defects) gathered in the microstructural and of the electrochemical cell's components [17][18]. As noted in **Fig. 3a**, the CDC-SPS sample exhibits a single semicircle that can be attributed to overlapped bulk and grain boundary contribution. Such kind of behavior is expected for a nanocrystalline ionic conductor where the large extension of the grain boundary is dominant over the minor bulk component. Nanoscaled ceria might include a very similar relaxation time for bulk and grain boundary due to the development of analogous length-scale of grain size and space charge region [19][20]. Hence, the separation of each contribution is unpractical and in this case, only total resistivity is considered. Conventionally sintered samples display two well-defined semicircles. According to the bricklayer theory, such responses are ascribed to a high and intermediate frequency related bulk and dopant-enriched grain boundary polarization [2]. Besides, the low-frequency semicircle in the CDC-Conv sample is semi-blocking, suggesting that the CaO facilitates the electrochemical diffusion mechanism of oxygen at the electrodes/electrolyte interface via electrochemical redox [21][22]. As evident in **Fig. 3.a**, CDC-Conv features relatively lower resistivity as of the GDC sample despite the macroscopic segregation of Ca at the grain boundary. Such an outcome can be attributed to the presence of a small g.b./bulk geometrical ratio as well as a low grain boundary ionic blocking factor ($\alpha_{g.b}$). The ion-blocking barrier generally arises from the crystalline disorder and other sources such as residual porosity, nanodomains, chemical segregation, etc. that eventually hinders the ionic migration pathway [23]. The measured $\alpha_{g.b}$ at 300 °C is about ~0.17 and ~0.9 for the CDC-Conv and GDC-Conv samples, respectively. As indicated in **Table 1**, the observed bulk relaxation frequency of the GDC-Conv sample is significantly larger (200 kHz) compared to the CDC-Conv sample (30 kHz), whereas a quite similar grain boundary relaxation frequency (~1 kHz) occurred. Even though not estimated, the presence of large resistivity in the CDC-SPS sample hints that the ion-blocking effect would be significantly high in this material. The temperature-dependent total electrical conductivity of the samples are plotted in an Arrhenius relationship in **Fig. 3.b**. As can be seen, the conductivity is minimum in the CDC-SPS sample throughout the examined temperatures. The low conductivity value in this nanostructured material can be ascribed to the presence of a high density of blocking barriers, strong intrinsic dopant-defect interaction as well as to crystalline disordered frozen in the material [24]. As registered, at low temperatures (< 350 °C) conductivity value of the CDC-Conv sample is larger than the GDC-Conv. Such results can be attributed to the fact that the ion-blocking factor is significantly smaller in CDC-Conv than GDC-Conv. On the other hand, at intermediate-high temperatures (> 400 °C), GDC-Conv illustrates a higher conductivity than the CDC-Conv due to the higher conductivity of $V_O^{\cdot\cdot}$ in the Gd-doped bulk. The activation energy value of the GDC sample (~0.9 eV) is reported a larger value compared to CDC materials (~0.65 eV), suggesting that the ionic migration mechanism/s at the CDC samples are easily activated and likely linked to fast diffusive properties of CaO [25].

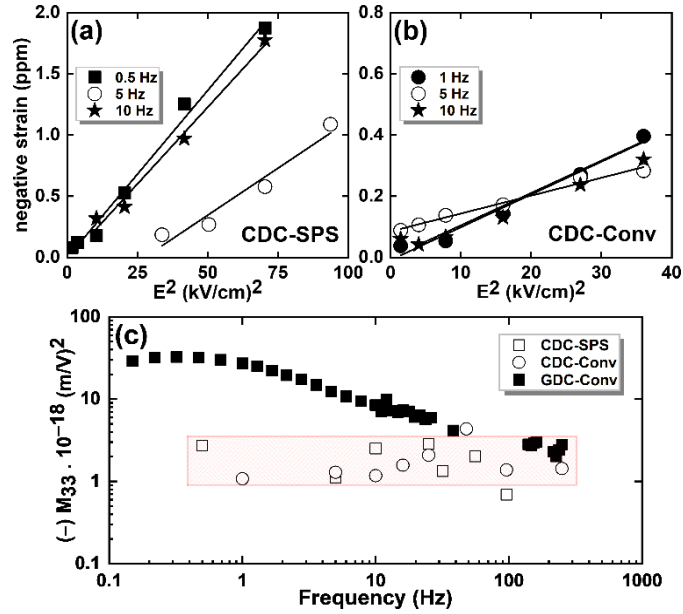

**Figure 4:** (a)-(b) The electrostrictive negative strain with the response to electric field square, at frequencies ranges between 0.15-10 Hz (c) The electrostriction strain coefficient ($M_{33}$) as a function of applied frequencies. The data of the GDC-Conv is taken after Ref. [6].

The electromechanical behavior of the CDC materials is highlighted in **Fig. 4**. As noticed, the CDC samples respond to the second harmonic of the applied frequency, confirming the physical characteristics of the electrostrictive response (strain ∝ $E^2$) in the materials. Within the measurement parameters (field and frequency), no strain saturation and frequency-dependent strain reduction are detected. The relaxation and saturation of electrostriction are common for GDC bulk, as shown in previous reports [5][6]. At certain electric field and frequency, the CDC-SPS is found to produce a slightly larger strain than the CDC-Conv sample (**Fig. 4.a** and **Fig 4.b**). The electrostrictive strain coefficient ($M_{33}$) with regard to the applied frequency is shown in **Fig. 4.c**. Over the measured frequency, both the CDC compound illustrate a very similar value of $M_{33}$ between ~1-3 x $10^{-18}$ (m/V)$^2$, even though having different microstructure, resistivity as well as ion blocking factor. However, such value is one order of magnitude smaller compared to GDC-Conv at the low-frequency regime, whereas the high-frequency value is comparable between all samples. As shown in our previous work, the low-frequency electrostriction coefficient in GDC material is strictly dependent on the ion-blocking barrier ($\alpha_{g.b}$) built in the materials [6]. Therefore, the absence of frequency-related $M_{33}$ relaxation suggests that the ion-blocking factor is not playing any dominant role in controlling electrostriction for CDC. Such a finding implies that the enhanced electrostatic and elastic dopant-defect interaction as induced by $Ca^{2+}$ can be another deciding factor in the electrostriction mechanism.

**Table 1**: A comparative analysis between samples, sintered in SPS and conventional method

| Sample ID | $f_{bulk}$ at 300 °C [kHz] | $f_{g.b.}$ at 300 °C [kHz] | $\alpha_{g.b.}$ at 300 °C | Conductivity at ~300 °C [$10^{-5}$ S/cm] | Electrostriction [$10^{-18}$ m$^2$/V$^2$] at 10 Hz |
|---|---|---|---|---|---|
| CDC-SPS | 20 | | | 1.4 | 2.5 |
| CDC-Conv | 30 | 0.5 | 0.17 | 3.7 | 1.2 |
| GDC-Conv | 200 | 1.5 | 0.9 | 3.1 | 8 |

In this work, highly dense nanostructured calcium doped ceria with a nominal composition of $Ce_{0.95}Ca_{0.05}O_{1.95}$ was fabricated by field-assisted spark plasma sintering technology (SPS) with a large ion-blocking effect. This results in a significantly lower electrical conductivity in comparison to the post-annealed sample. Moreover, both the CDC samples reveal fast ionic conductivity at low temperatures < 350 °C and an unconventional electromechanical response without any strain saturation and frequency relaxation effects. Remarkably, electrostriction in CDC follows no dependency on either grain size or ion-blocking effects rather directs by the presence of the strong electrostatic and elastic dopant-defect interaction $Ca^{2+}$-$V_O^{\cdot\cdot}$, leading to developing a steady $M_{33}$ from low to higher frequencies.

This research was supported by DFF-Research project grants from the Danish Council for Independent Research, Technology and Production Sciences, June 2016, grant number 48293 (GIANT-E) and European H2020-FETOPEN-2016-2017, project BioWings (Partially), grant number 801267.